\documentclass{PoS}

\title{Lattice artefacts in the Schr\"{o}dinger Functional coupling for strongly interacting theories }

\ShortTitle{Lattice artefacts in the SF coupling}

\author{\speaker{Stefan Sint and Pol Vilaseca}\\
        School of Mathematics, Trinity College Dublin, Dublin 2, Ireland\\
        E-mail: \email{sint@maths.tcd.ie, pol@maths.tcd.ie}}

\abstract{Models of Dynamical Electroweak Symmetry Breaking are expected to display a quasi-conformal
 scaling behaviour in order to accommodate experimental constraints. 
The scaling properties of a theory can be studied using finite volume renormalisation schemes. 
Among these, the most practical ones are based on the Schr\"{o}dinger Functional (SF). However, lattices accessible in
numerical simulations suffer from potentially large cutoff effects and special care has to be taken to remove these effects. 
Here we will study the standard setup of the SF with Wilson quarks and a setup with chirally rotated boundary conditions. 
We study the step scaling function for SU(2) and SU(3) gauge groups in the fundamental, 2-index symmetric and adjoint representations. 
We perform the O(a)-improvement of both setups to 1-loop order in perturbation theory, and we describe a way of 
minimising higher order cutoff effects by a redefinition of the renormalised coupling.}

\FullConference{The 30 International Symposium on Lattice Field Theory - Lattice 2012,\\
		June 24-29, 2012\\
		Cairns, Australia}

\begin{document}

\section{Introduction}

 Strongly interacting gauge theories other than QCD have been the focus of several studies in the 
recent past (see \cite{N2012_rev} for a review).
These can be constructed by coupling Yang-Mills fields to $N_{f}$ fermions 
transforming under higher dimensional representations $R$ of the gauge group $SU(N)$. For
certain values of $N$ and $N_{f}$ they are expected to show conformal or quasi conformal 
behaviour over some range of scales.

Schr\"{o}dinger Functional (SF) finite volume renormalisation schemes provide a useful tool for
studying non-perturbatively the scaling properties of the coupling. However, for the lattices accessible 
in numerical simulations these schemes
are known to display lattice artefacts arising from the bulk and from the temporal boundaries,
which may hide the universal continuum properties of the theory under consideration. Special care 
has to be taken in removing them such that continuum extrapolations can be done in a controlled
and reliable way.

In this work we examine the cutoff effects in the coupling at 1-loop order in perturbation theory, 
where the continuum limit is known. We will consider the standard SF setup with Wilson quarks
\cite{LNWW1992_lattSF,S1994_lattSF},
and the chirally rotated SF ($\chi$SF) which implements the mechanism of automatic $O(a)$
improvement \cite{S2011_lattSF}. Even after removing the $O(a)$ effects in the coupling, the remaining lattice artefacts are still large 
\cite{SV2011_lattSF,KTR2011_sfimp,KTR2012_sfimp} for the
typical lattice sizes used in simulations and further strategies have to be developed to minimise
them.

In the text we will briefly review some of the basic definitions of the SF coupling and the step
scaling function (SSF). We then study choices for the fermionic boundary conditions in the 
spatial directions which improve performance in numerical simulations. Finally, we propose a 
strategy to reduce the cutoff effects of the coupling in $SU(3)$ with a possible application                                                                                           
in $SU(2)$.

\section{The SF coupling and its cutoff effects.}

In SF schemes, a renormalised coupling constant is defined as the response of the system to variations of the background field (BF) $B$. Given
the effective action $\Gamma[B]$, the coupling and its perturbative expansion can be written as:
\begin{equation}
 \frac{1}{\overline{g}^{2}}=\frac{1}{k}\left.\frac{\partial\Gamma[B]}{\partial\eta}\right|_{\eta=0}; \qquad\qquad \overline{g}^{2}(L)\simeq g_{0}^{2}+p_{1}(L/a)g_{0}^{4}+O(g_{0}^{6}).
\label{g_2}
\end{equation}
with  $k=\left.\partial\Gamma_{0}[B]/\partial\eta\right|_{\eta=0}$, and $\Gamma_{0}[B]$ being the classical Wilson action for the BF.
The 1-loop coefficient $p_{1}=p_{1,0}+N_{f}p_{1,1}$ receives contributions from gauge and ghost fields ($p_{1,0}$) and from the fermion fields respectively ($p_{1,1}$).
The step scaling function (SSF) for a scale factor 2 and its expansion in perturbation theory are
\begin{equation}
 \Sigma(u,a/L)=\overline{g}^{2}(2L)|_{u=\overline{g}^{2}(L)}\stackrel{u\rightarrow0}{\sim} u+\Sigma_{1}(a/L)u^{2}+O(u^{3}),
\end{equation}
with a universal continuum limit
\begin{equation}
 \lim_{a\rightarrow0}\Sigma(u,a/L)=\sigma(u)\stackrel{u\rightarrow0}{\sim}u+\sigma_{1}u^{2}+O(u^{3}).
\end{equation}
The 1-loop term $\sigma_{1}=2b_{0}\ln(2)$ contains the $b_{0}$ coefficient of the $\beta$-function, and can also be separated into a pure gauge and a fermionic
part
\begin{equation}
 \sigma_{1}=\sigma_{1,0}+N_{f}\sigma_{1,1}; \qquad\qquad b_{0}=b_{0,0}+N_{f}b_{0,1}=\frac{1}{16\pi^{2}}\left(\frac{11N}{3}-N_{f}\frac{4}{3}T_{R}\right).
\end{equation}
The fermionic part $b_{0,1}$ depends on the representation $R$ under which the fermions transform via 
the normalisation constant $T_{R}$ ($T_{R}=1/2, N, (N+2)/2$ for fundamental, adjoint and symmetric fermions respectively).
Since the lattice SSF $\Sigma(u,a/L)$ depends on the details of the regularisation being used, we can define the relative deviations from
the pure gauge and fermionic continuum coefficients
\begin{equation}
 \delta_{1,0}(a/L)=\frac{\Sigma_{1,0}(a/L)-\sigma_{1,0}}{\sigma_{1,0}}, \qquad\qquad \delta_{1,1}(a/L)=\frac{\Sigma_{1,1}(a/L)-\sigma_{1,1}}{\sigma_{1,1}}, 
\end{equation}
and use them as a measure for the size of the cutoff effects of a specific regularisation. 

We consider abelian BF which arise by specifying the temporal boundary values for the gauge
fields to be 
\begin{equation}
 C_{k}=\frac{i}{L}\textrm{diag}\left(\phi_{1},\phi_{2},\phi_{3}\right)+\frac{i}{L}\eta\frac{\sqrt{3}}{2}\widetilde{\lambda}_{8}, \qquad\qquad C_{k}=\frac{i}{L}\textrm{diag}\left(\phi_{1},\phi_{2},\right)+\frac{i}{L}\eta\tau_{3}, \\
\label{C_k}
\end{equation}
\begin{equation}
 C'_{k}=\frac{i}{L}\textrm{diag}\left(\phi'_{1},\phi'_{2},\phi'_{3}\right)-\frac{i}{L}\eta\frac{\sqrt{3}}{2}\widetilde{\lambda}_{8}, \qquad\qquad  C'_{k}=\frac{i}{L}\textrm{diag}\left(\phi'_{1},\phi'_{2},\right)-\frac{i}{L}\eta\tau_{3}, 
\label{C_kp}
\end{equation}
for $SU(3)$ and $SU(2)$ respectively. 

The basis for the $SU(3)$ generators is such that $\widetilde{\lambda}_{8}=\frac{1}{\sqrt{3}}\textrm{diag}(2,-1,-1)$,
and $\tau_{3}$ is the third Pauli matrix. The values for the phases $\phi_{i}$ and $\phi'_{i}$ are collected in table \ref{phases}.
 Note that the term with $\eta$ is usually
included in the definition of the phases $\phi_{i}$. We choose to write it explicitly for later
convenience. This term can be thought of as a deformation of the BF in the direction in the algebra $su(3)$ ($su(2)$) given
by the generator $\widetilde{\lambda}_{8}$ ($\tau_{3}$). It is used to define the renormalised coupling eq.(\ref{g_2})
and is fixed to zero after differentiation.
\begin{table}[ht]
 \centering
 \begin{tabular}{c c c c c c c}
    \hline\hline
              & $\phi_{1}$ & $\phi_{2}$ & $\phi_{3}$ & $\phi'_{1}$ & $\phi'_{2}$ & $\phi'_{3}$  \\
    \hline
    $SU(3)$ & $-\pi/3$ & $0$ & $\pi/3$ & $-\pi$ & $\pi/3$ & $2\pi/3$   \\
    $SU(2)$ & $-\pi/4$ & $\pi/4$ & $-$ & $-3\pi/4$ & $3\pi/4$ & $-$  \\
 \end{tabular}
 \caption{Phases of the boundary fields for $SU(3)$ \cite{LSWW1994} and $SU(2)$ \cite{LNWW1992_lattSF} used in this work.}
\label{phases}
\end{table}
The spatial boundary conditions for the gauge fields are periodic. Fermion fields
satisfy temporal boundary conditions with projectors (\cite{S1994_lattSF,S2011_lattSF}).
The spatial boundary conditions for 
the fermions are discussed in section 3.

Lattice artefacts can be large for the lattices accessible in numerical simulations.
One must therefore find regularisations in which the cutoff effects are as small as possible even for 
small lattices.
$O(a)$ effects can be cancelled following Symanzik's improvement programme by adding a set of counterterms to the action. In the SF setup 
$O(a)$ improvement is achieved by adding the clover term with coefficient $c_{\rm sw}$ to cancel effects coming from the bulk, and a set of 
boundary counterterms  with coefficients $c_{\rm t}$, $\widetilde{c}_{\rm t}$, $c_{\rm s}$ and $\widetilde{c}_{\rm s}$ to remove the $O(a)$ effects arising
from the boundaries. For our specific choice of BF and boundary conditions, only $c_{\rm t}$ and $\widetilde{c}_{\rm t}$ are needed.
1-loop $O(a)$ improvement of the coupling is implemented by setting the tree level coefficients to $c_{\rm t}^{(0)}=1$, 
$\widetilde{c}_{\rm t}^{(0)}=0$ and $c_{\rm sw}^{(0)}=1$, and by fixing the coefficient $c_{\rm t}^{(1)}$ to its correct 1-loop value.
The bare mass is set to $m_{0}=0$ throughout. 

In the $\chi$SF setup the bulk is automatically free from $O(a)$ effects once the coefficient $z_{\rm f}$ of a dimension 3 boundary counterterm
has been tuned \cite{S2011_lattSF}. $O(a)$ effects coming from the boundaries are removed by fixing 2 coefficients $c_{\rm t}$
and $d_{\rm s}$. As for the SF setup, 1-loop improvement in the coupling 
is achieved by setting the coefficients to their tree level values $z_{\rm f}=1$, 
$d_{\rm s}^{(0)}=0.5$ and $c_{\rm t}^{(0)}=1$, and by setting only $c_{\rm t}^{(1)}$ to the correct 1-loop value.
The bare mass is set to $m_{0}=0$ as in the SF.

The boundary counterterm $c_{\rm t}^{(1)}$ has a gauge and fermion parts ($c_{\rm t}^{(1,0)}$ and $c_{\rm t}^{(1,1)}$ respectivelly).
$c_{\rm t}^{(1,1)}$ for the SF setup with $c_{\rm sw}=1$ was calculated in \cite{SS1996}.
For the $\chi$SF we consider a setup with no clover term
($c_{\rm sw}=0$) and a setup with $c_{\rm sw}=1$. For the fundamental representation the values of $c_{\rm t}^{(1,1)}$ are 
the same in $SU(2)$ and $SU(3)$, and for the different setups they read
\begin{equation}
 \left.c_{\rm t}^{(1,1)}\right|_{SF}=0.0191405(2),\quad \left.c_{\rm t}^{(1,1)}\right|_{\chi SF,c_{\rm sw}=0}=-0.00661445(5), \quad \left.c_{\rm t}^{(1,1)}\right|_{\chi SF,c_{\rm sw}=1}=0.006888(3). 
\end{equation}
The values of $c_{\rm t}^{(1,1)}$ for a representation $R$ can be obtained as observed in \cite{Karavirta:2011mv}
by scaling the value for the fundamental representation according to
\begin{equation}
 c_{\rm t}^{(1,1)}(R)=2T(R)c_{\rm t}^{(1,1)}(F).
\end{equation}

\section{Spatial boundary conditions and the condition number}

\begin{figure}[h!]
\begin{center}\includegraphics[clip=true,scale=0.47]{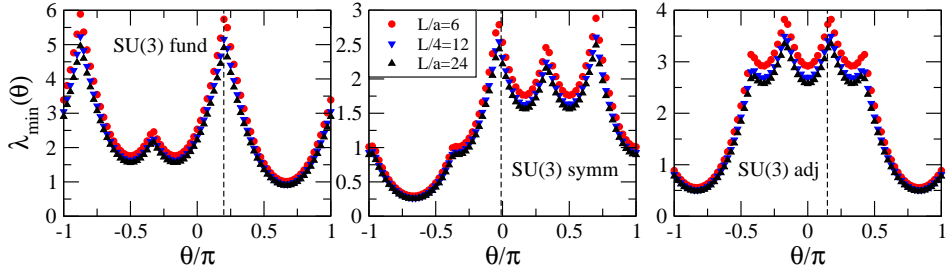}
\end{center}
\caption{Lowest eigenvalue $\lambda_{min}$ (in units of $L^{-2}$) as a function of $\theta$
 for the fundamental and symmetric representations of $SU(3)$. The vertical discontinuous
line denotes the selected value of $\theta$. The plots for the representations of 
$SU(2)$ are not shown.} 
\label{eigenv}
\end{figure}

The spatial boundary conditions for the fermion fields are taken to be periodic up to a phase
\begin{equation}
 \psi(x+L\hat{k})=e^{i\theta/L}\psi(x);\qquad\qquad\overline{\psi}(x+L\hat{k})=\overline{\psi}(x)e^{-i\theta/L}; \qquad\qquad k=1,2,3.
\end{equation}
%Although the value of $\theta$ is of no fundamental importance, it is part of the definition of the renormalization scheme and
The value of $\theta$ 
can be chosen following the guideline principle established in \cite{SS1996}. In numerical simulations the 
efficiency of the inversion algorithms depends on the condition number, i.e. the ratio $\lambda_{max}/\lambda_{min}$ between 
the largest and smallest eigenvalues of the squared fermion matrix.%In numerical simulations the efficiency of the inversion%
%algorithms de depends on the condition number $\kappa=(\lambda_{N}/\lambda_{0})^{1/2}$, $\lambda_{N}$ and $\lambda_{0}$ being the largest and smallest eigenvalues%
%of the squared fermion matrix $\Delta_{2}$ respectively%
While $\lambda_{min}$ strongly depends on $\theta$, $\lambda_{max}$ is rather insensitive to it.
Hence, by chooding the angle $\theta$ such that $\lambda_{min}$ in the BF is as large as possible the condition number is minimised,
optimising the performance of the simulations in the perturbative regime.
The lowest eigenvalue as a a function of $\theta$ is shown in figure \ref{eigenv}. The profile $\lambda_{min}(\theta)$ depends on the chosen BF and the fermion representation considered.
Near optimal values of $\theta$ for the groups and representations studied in this work can be found in table \ref{angle} and will be kept for the rest of the study.
\begin{table}[ht]
 \centering
 \begin{tabular}{c c c c c c}
              & $SU(2)_{f}$ & $SU(2)_{a}$ & $SU(3)_{f}$ & $SU(3)_{a}$ & $SU(3)_{s}$  \\
    \hline
    $\theta$  & $0$ & $2\pi/5$ & $\pi/5$ & $\pi/6$ & $0$ \\
  \end{tabular}
 \caption{Angle $\theta$ in the fermionic boundary conditions in the spatial directions which leads to a highest $\lambda_{min}$.}
\label{angle}
\end{table}

\section{Cutoff effects}

\begin{figure}[h!]
\begin{center}\includegraphics[clip=true,scale=0.48]{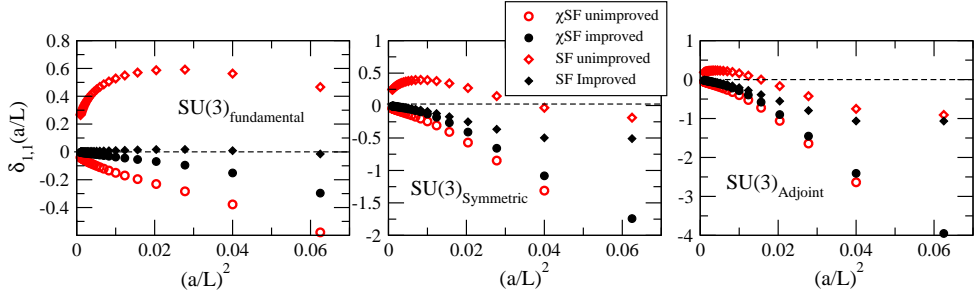}
\end{center}
\caption{Cutoff effects for the fermionic part of the SSF for $SU(3)$ in the fundamental, symmetric and adjoint representations. Both setups
(SF and $\chi$SF) are shown before and after $O(a)$ improvement.} 
\label{d11}
\end{figure}

After fixing the improvement coefficients to their correct values we see that cutoff 
effects in the fundamental representations of $SU(2)$ and $SU(3)$ are essentially zero for 
the SF and very small for the $\chi$SF (see figure \ref{d11} and references \cite{SV2011_lattSF,KTR2011_sfimp,KTR2012_sfimp}). The situation is drastically 
different when considering
other fermion representations. There the cutoff effects in the SSF remain large even after
$O(a)$ improvement. 
This was already observed
in \cite{SV2011_lattSF} where a possible cure was prescribed by making the BF weaker. A more
systematic study has been done in \cite{KTR2011_sfimp,KTR2012_sfimp} 
exploring different modifications to the BF.
However, the price to pay when abandoning the original choices of BF is a loss 
in the statistical behaviour of the signals \cite{KTR2012_sfimp}.
Here we choose to follow a different strategy for reducing the higher order cutoff effects,
in which the BF is left intact and the coupling constant is modified.

\subsection{Strategy for SU(3)}

In order to define the coupling constant in $SU(3)$ one takes derivatives of the 
effective action $\Gamma$ with respect to the parameter $\eta$ in eq.(\ref{C_k}). 
In $SU(3)$ another renormalised observable $\overline{\textrm{v}}$ \cite{LSWW1994} can be defined and added to the
coupling such that one obtains a family of renormalised couplings
\begin{equation}
 \frac{1}{\overline{g}_{\nu}^{2}(L)}=\frac{1}{\overline{g}^{2}(L)}-\nu\overline{\textrm{v}}(L); \qquad\qquad \overline{\textrm{v}}(L)=\frac{1}{k}\frac{\partial}{\partial\nu}\left.\left(\left.\frac{\partial\Gamma}{\partial\eta}\right|_{\eta=0}\right)\right|_{\nu=0}.
\label{g_nu}
\end{equation}
parameterised by the real number $\nu$.
$\overline{\textrm{v}}(L)$ does not contribute at tree level and is thus a pure
quantum effect.
The coupling $\overline{g}_{\nu}^{2}$ can be understood as the derivative of the effective action along the direction in 
the algebra $su(3)$ given by $(\frac{\sqrt{3}}{{2}}\widetilde{\lambda}_{8}+\nu \widetilde{\lambda}_{3})$, with $\widetilde{\lambda}_{3}=\textrm{diag}(0,1,-1)$.
This can be seen by replacing $\eta \frac{\sqrt{3}}{{2}}\widetilde{\lambda}_{8}\longrightarrow\eta(\frac{\sqrt{3}}{{2}}\widetilde{\lambda}_{8}+\nu \widetilde{\lambda}_{3})$
in eqs. (\ref{C_k}) and (\ref{C_kp}). Since $\eta$ is later fixed to $0$, this change does not affect the BF at all.

\begin{figure}[h!]
\begin{center}\includegraphics[clip=true,scale=0.47]{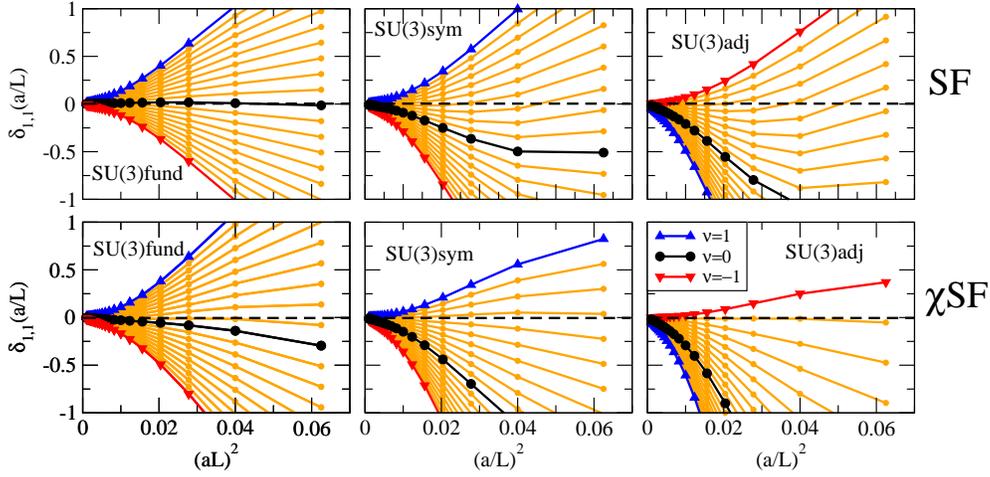}
\end{center}
\caption{Cutoff effects on the SSF in $SU(3)$ for several choices of $\nu$. Black dotted lines correspond to the original choice $\nu=0$.
Neighbouring lines are separated by an interval $\Delta\nu=0.1$.} 
\label{delta_nu}
\end{figure}

By modifying the value of $\nu$ we are able to explore the cutoff effects associated to the 
different couplings $\overline{g}_{\nu}^{2}(L)$ for the fundamental, symmetric and adjoint
representations of $SU(3)$ (see figure \ref{delta_nu}).
In all the cases we consider there is a strong dependence of the cutoff effects
with $\nu$ and one can always find a value of $\nu$ for which they are minimal. For the fundamental
representation in the standard SF $\nu=0$ minimises the cutoff effects. However, by moving to larger 
values of $\nu$ the higher order effects rapidly increase towards a situation closer to what was 
found initially for the adjoint or symmetric representations. The good behaviour of the fundamental
representation seems to be an accident of having $\nu=0$ as an initial choice. This value
is not universal and depends on the regularisation considered. For instance, for the $\chi$SF the
optimal value is $\approx0.1$. %It would be interesting to know the optimal value of $\nu$ 
%for other regularisations such as staggered fermions.
%For the symmetric representation, the higher order effects disappear for $\nu\simeq0.4$ in the SF
%and $\nu\simeq0.7$ in the $\chi$SF. The adjoint representation displays
%the same kind of dependence on $\nu$ and the cutoff effects are minimised for $\nu\simeq-0.6$
%for the SF and $\nu\simeq-0.9$ for the $\chi$SF.

\subsection{A possible strategy for SU(2)}

While the 1-loop cutoff effects in the SSF for $SU(3)$ seem well under control when including the observable $\overline{\textrm{v}}$,
it remains an open question wether the lattice artefacts for $SU(2)$ can be reduced in a similar fashion. In
$SU(2)$ there is no generator other than $\tau_{3}$
giving abelian directions to build a family of couplings as in eq.(\ref{g_nu}). However, one might try to 
extend the same argument to include all other generators $\tau_{i}$ of the algebra to define a family
of observables
\begin{equation}
 \eta\tau_{3}\longrightarrow\eta\left(\tau_{3}+\sum_{i=1}^{2}\nu_{i}\tau_{i}\right);\qquad\qquad \overline{\textrm{v}}_{i}(L)=\frac{1}{k}\frac{\partial}{\partial\nu_{i}}\left.\left(\left.\frac{\partial\Gamma}{\partial\eta}\right|_{\eta=0}\right)\right|_{\nu_{i}=0}.
\end{equation}
The addition of the terms $\nu_i\tau_{i}$ is a way of defining the observables $\overline{v}_{i}$.
As long as $\eta$ is set to $0$ in the final calculation the resulting BF will still be abelian.
It can be shown that the $\overline{v}_{i}$ do not contribute at tree level and can be used
to construct a new family of couplings by taking derivatives along a direction $\mathbf{V}$ in
 the algebra
\begin{equation}
  \frac{1}{\overline{g}_{\mathbf{V}}^{2}(L)}=\frac{1}{\overline{g}^{2}(L)}-\sum_{i=1}^{2}\nu_{i}\overline{\textrm{v}}_{i}(L)
\end{equation}
A similar definition can be given for $SU(3)$ by considering all 8 generators of the algebra. Whether
these observables will help or not in reducing the cutoff effects in $SU(2)$ is currently under
investigation.
\section{Conclusions}
We have shown a way of controlling the 1-loop cutoff effects in the SSF for the representations of 
$SU(3)$ without modifying the BF. The results of this perturbative calculation are encouraging, but
it remains to be checked whether the cutoff effects at higher orders in perturbation theory can be removed
in a similar fashion. Moreover, experience should be accumulated with the non-perturbative behaviour of $\overline{\textrm{v}}$
for representations other than the fundamental. The
original choice of $\nu=0$ for $SU(3)$ was taken in order to 
obtain the best signal to noise ratio for the coupling in the pure gauge theory \cite{LSWW1994}, so for the theories we are 
considering one probably has to find a balance between 
the reduction of cutoff effects and the quality of the signal.

\section*{Acknowledgements}
The authors gratefully acknowledge support by SFI under grant 11/RFP/PHY3218 and by the
EU under grant agreement number PITN-GA-2009-238353 (ITN STRONGnet).

\end{document}